\documentclass[aps,twocolumn,superscriptaddress]{revtex4}
\usepackage{graphicx}

\begin{document}
\title{Superconducting properties of MgB$_2$ probed by radiation-induced
disordering}
\author{A.~E.~Karkin}
\email[E-mail: ]{karkin@orar.zar.ru}
\affiliation{Institute of Metal Physics UB RAS, Ekaterinburg 620219,
Russia}
\author{V.~I.~Voronin}
\affiliation{Institute of Metal Physics UB RAS, Ekaterinburg 620219,
Russia}
\author{T.~V.~Dyachkova}
\affiliation{Institute of Solid State Chemistry UB RAS, Ekaterinburg,
Russia}
\author{A.~P.~Tyutyunnik}
\affiliation{Institute of Solid State Chemistry UB RAS, Ekaterinburg,
Russia}
\author{V.~G.~Zubkov}
\affiliation{Institute of Solid State Chemistry UB RAS, Ekaterinburg,
Russia}
\author{Yu.~G.~Zainulin}
\affiliation{Institute of Solid State Chemistry UB RAS, Ekaterinburg,
Russia}
\author{B.~N.~Goshchitskii}
\email[E-mail: ]{bng@imp.uran.ru}
\affiliation{Institute of Metal Physics UB RAS, Ekaterinburg 620219,
Russia}
\date{\today}

\begin{abstract}
The electrical resistivity $\rho$ and upper critical field $H_{c2}$ were
measured in MgB$_2$ disordered by nuclear reactor neutrons. We found
that superconducting temperature $T_c$ decreases under irradiation from
40 to 5 K. Despite strong disordering (more than 10 dpa (displacements
per atom)), the initial crystal structure is preserved. The residual
resistivity $\rho_0$ increases from 0.35 to 2 m$\Omega$ cm
while $-dH_{c2}/dT$ remains approximately unchanged
($-dH_{c2}/dT \approx 0.5$ T/K), and such behavior may be
interpreted as decrease in the density of electronic states.
\end{abstract}
\maketitle

With the discovery of superconductivity (SC) in binary compound MgB$_2$
with relatively high $T_c \approx 40$ K \cite{Akimitsu}, a question
arises about the possible origin of SC in such a simple compound. It is
well known that high density of electronic states at the Fermi level
$N(E_{\text{F}})$, similar to soft phonon frequencies, can lead to enhancement
of $T_c$ in BCS phonon-mediate superconductivity \cite{McMillan}, while
some form of exotic coupling can take place in this case. One of the ways
to determine the nature of coupling is to study the response of SC on
disordering induced by irradiation with high-energy particles.

Disordering of intermetallic compounds, such as A15 (Nb$_3$Sn, V$_3$Si),
with high $N(E_{\text{F}})$ leads to decrease of $T_c$ from  15 - 20 K to 1 - 3 K
due to drop in $N(E_{\text{F}})$ at disordering \cite{Karkin}. In low-$T_c$
compounds (Mo$_3$Si and Mo$_3$Ge), irradiation leads to increase of $T_c$
(from 1.5 to 7 K) due to rise in $N(E_{\text{F}})$ as well as due to softening of
phonon frequencies. It means that individual features of the electronic
structure disappear under disordering, and the superconducting properties
of the disordered compounds become similar to the properties of amorphous
materials obtained by various methods.

The influence of disordering on SC of high-$T_c$ compounds is not similar
to that of A15: irradiation leads to very fast degradation of SC
($T_c = 0$) \cite{Aleksashin}. Although the origin of $T_c$-degradation
is not clear (because the mechanism of SC in high-$T_c$ is unknown),
there is a significant difference between intermetallic compounds and
more exotic materials in the response to disordering.

\begin{figure}
\includegraphics[keepaspectratio,width=2.8in]{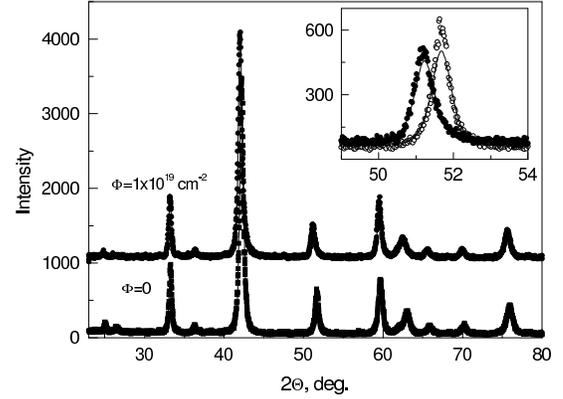}
\caption{X-ray diffraction patterns for initial ($\Phi = 0$) and
irradiated ($\Phi = 1 \cdot 10^{19}$ cm$^{-2}$) MgB$_2$ samples.}
\label{fig:1}
\end{figure}

\begin{figure}
\includegraphics[keepaspectratio,width=2.8in]{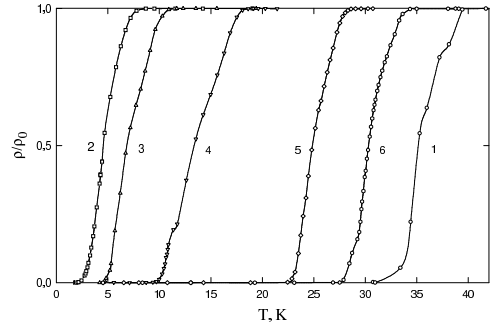}
\caption{Resistivity superconducting transition curves for initial (1),
irradiated (2) and annealed at 200, 300, 400, 500$^\circ$C (3-6)
MgB$_2$ samples.}
\label{fig:2}
\end{figure}

\begin{figure}
\includegraphics[keepaspectratio,width=2.8in]{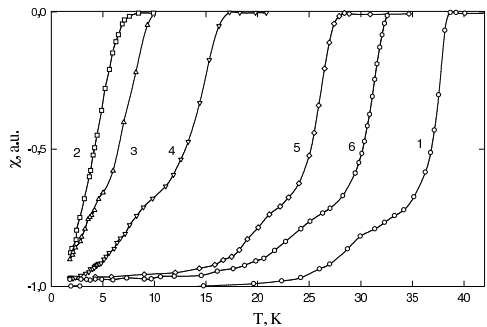}
\caption{AC-susceptibility superconducting transition curves for
initial (1),
irradiated (2) and annealed at 200, 300, 400, 500$^\circ$C (3-6)
MgB$_2$ samples.}
\label{fig:3}
\end{figure}

In this paper we studied the transport properties of polycrystalline
samples ($0.05\cdot 1\cdot5$ mm$^3$) irradiated in nuclear reactor IVV-2M
at $T \approx 350$ K up to fluences of thermal neutrons
$\Phi = 1\cdot 10^{19}$ cm$^{-2}$, and of fast neutrons, to
$5 \cdot 10^{18}$ cm$^{-2}$ (to the total dose - of more than 10 dpa).
The MgB$_2$ powder was prepared by the technology described
in Ref.~\cite{Gerashenko}. The samples were compacted under 9 Gpa at room
temperature without any subsequent heat treatment. Such sample preparation
produces a stressed structure, which results in broadening of X-ray peaks
(Fig.~\ref{fig:1}) and resistivity $\rho$ (Fig.~\ref{fig:2}) and
ac-susceptibility $\chi$ (Fig.~\ref{fig:3}) superconducting transitions.
However the initial crystal structure (C32) is preserved under irradiation
inducing such high damage to the material.

Fig.~\ref{fig:1} shows X-ray diffraction patterns (Cu-K$_\alpha$)
for the initial
and the irradiated samples of MgB$_2$. The samples contain traces
($\sim$ 3 \%) of MgO. The structure parameters were refined with Rietveld
analysis. Some discrepancy between the measured and the calculated
curves (see the insert in Fig.~\ref{fig:3} ) was connected with the
sample texture. Irradiation leads to anisotropic expansion of the crystal
lattice (the cell volume increase is about 1.4 \%) with faster increase
of c, as is well seen in the insert in Fig.~\ref{fig:3} and
Table~\ref{table:1}. The broadened diffraction peaks in the initial
samples become narrower after irradiation, which is an evidence of partial
relief of stress. The Rietveld analysis refinement yields some decrease in
the Mg site occupation numbers (if B site occupation is fixed equal to
unit), although we do not expect appearance of anti-structural defects
because of the difference in the atomic sizes of Mg and B.
\begin{table}[h]
\caption{Lattice parameters $a$, $c$, cell volume $V$  and site
occupation number (Mg) for initial and irradiated  MgB$_2$ samples.}
\label{table:1}
\begin{tabular*}{\hsize}{l@{\extracolsep{0ptplus1fil}}l@{\extracolsep{0ptplus1fil}}l@{\extracolsep{0ptplus1fil}}}
\toprule
&Initial
&Irradiated\\
\colrule
$a$, \AA
&3.0878(2)
&3.0953(3)\\
$c$, \AA
&3.5216(4)
&3.5533(4)\\
$V$, \AA$^3$
&29.080(4)
&29.482(5)\\
$c/a$
&1.140
&1.148 \\
$N$(Mg)
&0.96(4)
&0.89(5)\\
\botrule
\end{tabular*}
\end{table}

\begin{figure}
\includegraphics[keepaspectratio,width=2.8in]{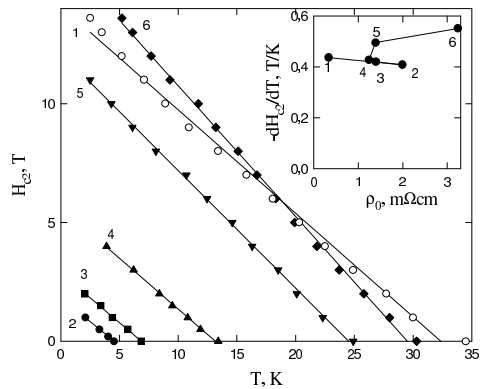}
\caption{Upper critical field 
$H_{c2}$ vs. temperature $T$ for initial (1),
irradiated (2) and annealed at 200, 300, 400, 500$^\circ$C (3-6)
MgB$_2$ samples.
Insert shows $-dH_{c2}/dT$ as a function of
normal state resistivity $\rho_0$.}
\label{fig:4}
\end{figure}

\begin{figure}
\includegraphics[keepaspectratio,width=2.8in]{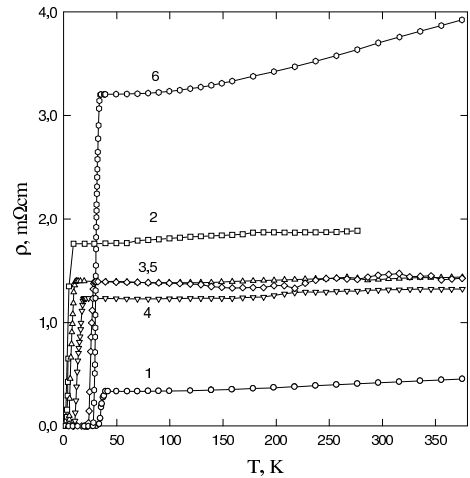}
\caption{Temperature dependences of  resistivity $\rho$ for initial(1),
irradiated(2) and annealed at 200, 300, 400, 500$^\circ$C(3-6)
MgB$_2$ samples}
\label{fig:5}
\end{figure}

Figs.~\ref{fig:2}, \ref{fig:3} show transformations of $\rho$- and
$\chi$-transitions in irradiated and annealed samples. There are no
essential changes of transition widths under irradiation, which testifies
to homogeneous distribution of radiation defects. Annealing at 500 K
recovers 85 \% of $T_c$. The values of the upper critical field $H_{c2}$
determined by half-width of the normal state resistivity show almost the
same slope for the initial, irradiated and post-annealed samples
(Fig.~\ref{fig:4}). $H' = -dH_{c2}/dT$ shows low dependence on
normal state resistivity $\rho_0$ (see the insert in Fig.~\ref{fig:4}).
The resistivity $\rho(T)$ increases after irradiation, decreases after
annealing up to 300$^\circ$ C and grows again at subsequent annealing
reaching 4 m$\Omega$ cm (Fig.~\ref{fig:5}). We consider that such a
behavior of  $\rho(T)$ may be partly due to intergrain transport,
therefore $\rho(T)$ is not completely an intrinsic property. As is
expected, $H'$ within the dirty limit should increase with the decrease
in the mean free path of electron under disordering, therefore the small
change in $H'$ may be interpreted as the decrease of $N(E_{\text{F}})$ with
decreasing $T_c$. It may be supposed that, similar to A15 compounds,
there is a fine structure in $N(E)$ in the vicinity of the Fermi level,
which is smeared under radiation-induced disordering.

Work supported by  the Russian State contracts Nos 107-1(00)-P,
107-19(00)-P, 108-31(00)-P, the Russian State Program of Support to
Leading Scientific Schools (Project No. 00-15-96581) and RFBR Grant
No. 00-02-16877.

\end{document}